\documentclass[journal]{IEEEtran}

\usepackage{amsmath}
\usepackage{graphicx}
\usepackage{float}
\usepackage[shortlabels]{enumitem}
\usepackage[dvipsnames]{xcolor}

\begin{document}

\title{Addressing Bias Early in the AI Lifecycle: Social Determinants of Health in Functional and Cognitive Assessment Labels of Nursing Home Residents}

\author{Isabel Amaya-Rodr\'iguez$^{a,b,c,d}$,
        Cristina Martin$^{b,c,e}$,
        Nekane Larburu$^{b,c}$,
        Teresa Garcia-Navarro$^{b}$,      
        
        Roberto Gonz\'alez Novas$^{f}$,
        Alberto Rodr\'iguez Taboada$^{f}$,
        Maite Garc\'ia-Sebastian$^{g}$,
        Ainara Estanga$^{g}$,
        Mirian Ecay-Torres$^{g}$,
        Pablo Mart\'inez-Lage$^{g}$,
        Miguel A. Armengol de la Hoz$^{h,i}$,
        Cristina Rubio-Escudero$^{a}$,
        and Iv\'an Mac\'ia$^{c}$%
\thanks{$^{a}$Computer Languages and Systems, University of Seville, Seville, Spain.
$^{b}$Fundaci\'on Vicomtech, Basque Research and Technology Alliance (BRTA), Mikeletegi 57, 20009 Donostia-San Sebasti\'an, Spain.
$^{c}$eHealth Group, Bioengineering Area, Biodonostia Health Research Institute, 20014 Donostia-San Sebasti\'an, Spain.
$^{d}$Institute of Biomedicine of Seville (IBiS), Seville, Spain.
$^{e}$Faculty of Engineering, University of Deusto, Avda. Universidades 24, 48007 Bilbao, Spain.
$^{f}$DomusVi, Spain.
$^{g}$Center for Research and Memory Clinic, CITA-Alzheimer Foundation, 20009 Donostia-San Sebasti\'an, Spain.
$^{h}$Data Science Lab, Regional Ministry of Health of Southern Spain, Seville, Spain.
$^{i}$Pharmacotherapy Follow-up and Personalised Medicine, Institute of Biomedicine of Seville (IBiS), Seville, Spain.
Corresponding author: I. Amaya-Rodr\'iguez (e-mail: iamaya@alum.us.es).}}

\markboth{Amaya-Rodr\'iguez \MakeLowercase{\textit{et al.}}: Social Determinants of Health in Functional and Cognitive Assessment Labels}{}

\maketitle

\begin{abstract}
Fairness in clinical artificial intelligence (AI) is usually evaluated after model training, yet bias can enter earlier in the AI lifecycle, during data collection and labelling. When routine clinical assessments serve as target variables, systematic differences in these labels across sociodemographic groups are carried into model development, validation and updating. This study addresses this early lifecycle stage by auditing labels from three standardised instruments used in residential care: the Barthel Index (BI), the Mini-Mental State Examination (MMSE) and the Global Deterioration Scale (GDS). Using electronic health records of 2,407 residents from 12 nursing homes in the Basque Country, Spain (2011--2021), we assessed four social determinants of health (SDoH)---sex, educational attainment, lifetime work activity and marital status---with two group fairness metrics, Disparate Impact (DI) and Statistical Parity Difference (SPD), computed for each outcome category with bootstrap confidence intervals. Disparities were found for all four attributes and across the three instruments: men, residents with intermediate education and never-partnered residents were more frequently classified in categories indicating preserved function or cognition. These disparities may reflect genuine differences in health status, measurement properties of the instruments, assessment practices or selective admission to residential care, and cannot be attributed to a single cause with the present design. Regardless of their origin, any model trained on these labels would inherit them, and they could persist through later lifecycle stages unless explicitly examined. These findings support pre-training label auditing as the first step of a continuous fairness assessment spanning model development, validation, deployment and updating.
\end{abstract}

\begin{IEEEkeywords}
Algorithmic Fairness, Label Bias, Social Determinants of Health, AI Lifecycle, Residential Care, Geriatric Assessment
\end{IEEEkeywords}

\section{Introduction}

\IEEEPARstart{H}{e}alth is a complex system in which multiple biological, behavioural, social and environmental factors interact throughout the life course. Beyond genetic and biological determinants, health trajectories are shaped by human factors and social determinants of health (SDoH), including educational attainment, occupational history, socioeconomic conditions, social relationships and access to healthcare \cite{majoka2021sdoh}. These factors accumulate over time and may influence not only health status in later life, but also functional capacity, cognitive performance and the way health conditions are assessed and documented \cite{livingston2020dementia}.

Population ageing has increased the number of older adults requiring long-term care, with residential care facilities increasingly functioning as complex health and social care environments rather than exclusively medical settings. Residents receive not only medical and nursing care, but also assistance with activities of daily living (ADLs), rehabilitation, occupational and cognitive stimulation, and social and emotional support \cite{fine2005dependence}. Functional independence in ADLs is therefore a central dimension of healthy ageing and an important determinant of care needs, resource allocation and quality of life. In this context, cognitive impairment and dementia, particularly Alzheimer's disease and related disorders, represent some of the most challenging conditions in residential care because of their close relationship with functional decline, dependency and increasing care requirements \cite{livingston2020dementia}.

The complexity of these settings is further increased by the heterogeneity of both residents and care processes. Residents from different generations may share the same facility while having markedly different educational backgrounds, occupational histories, family structures, socioeconomic circumstances and cultural experiences. These life-course differences may shape cognitive reserve, functional capacity and patterns of cognitive ageing \cite{feldberg2024occupation,arora2024education}. At the same time, assessment and care involve multidisciplinary teams---including physicians, nurses, psychologists, physiotherapists, occupational therapists and care assistants---who may observe and document different aspects of residents' functioning. Consequently, variability may arise not only from differences between individuals, but also from the context and professionals involved in their assessment. Such heterogeneity does not necessarily constitute bias; however, when systematic differences in assessment or documentation are associated with sociodemographic characteristics and are unrelated to the underlying clinical condition, they may become a relevant source of inequity \cite{rajkomar2018fairness,challen2019bias}.

The assessment of cognitive and functional status in residential care commonly relies on standardised clinical instruments. The Barthel Index (BI) evaluates independence in basic ADLs, whereas the Mini-Mental State Examination (MMSE) and the Global Deterioration Scale (GDS) provide complementary measures of cognitive status and cognitive deterioration \cite{mahoney1965barthel,folstein1975minimental,reisberg1982gds}. These instruments are widely used and clinically established. However, they provide a structured assessment at a particular point in time and may be influenced by characteristics such as educational attainment, language, background or the circumstances under which the assessment is performed; for instance, MMSE performance varies markedly with age and educational level, and education-adjusted cut-off points have been proposed for the Spanish population \cite{crum1993,lobo1999}. Educational attainment, for example, has been consistently associated with cognitive performance and cognitive reserve, while occupational complexity and intellectual activity throughout the life course may also influence later-life cognitive functioning \cite{feldberg2024occupation,arora2024education,lojo2012actividad}. Similarly, sex and gender differences in life-course experiences may contribute to differences in cognitive ageing and dementia risk \cite{mielke2022sexgender}.

These considerations have encouraged the exploration of complementary sources of longitudinal information. Wearable devices and other digital technologies can continuously capture patterns of physical activity, mobility, sleep and behaviour, potentially providing information that is not available through periodic clinical assessments \cite{cejudo2025smart}. Artificial intelligence (AI) and machine learning techniques can further integrate these heterogeneous data sources to identify complex patterns and generate individualised risk estimates. However, the clinical value of these systems depends not only on predictive performance but also on the quality, representativeness and fairness of the data used to develop them \cite{rajkomar2018fairness,challen2019bias}.

This issue becomes particularly important when clinical assessments are subsequently used as target variables for AI models. If outcome labels systematically differ across sociodemographic groups for reasons unrelated to the underlying clinical condition, a model trained on these labels may learn and reproduce these disparities rather than solely reflecting clinically relevant differences \cite{rajkomar2018fairness,challen2019bias}. This form of label choice bias has been shown to produce substantial inequities in deployed clinical algorithms \cite{obermeyer2019}, and corresponds to the measurement and historical biases that lifecycle frameworks locate at the data collection and labelling stages \cite{suresh2021}. Biases in healthcare data can arise at multiple stages, including differences in access to care, clinical documentation, measurement and assessment processes \cite{challen2019bias}. Missing or systematically incomplete information may further contribute to inequitable model performance \cite{getzen2023missing}. Thus, potential sources of bias should be considered not only at the model-development stage, but also in the clinical data and in the assessment-derived labels BI, MMSE and GDS classifications used as the basis for training

SDoH may therefore represent both determinants of health and sources of heterogeneity in clinical data. Educational attainment and occupational trajectories are associated with cognitive reserve and later-life cognitive performance \cite{feldberg2024occupation,arora2024education,lojo2012actividad}. Likewise, marital status and social relationships have been associated with cognitive impairment and dementia risk, although these associations may reflect broader differences in social support, physical health and life-course circumstances \cite{liu2019marital,sommerlad2018marriage}. These factors are particularly relevant in residential care, where clinical outcomes are interpreted within highly heterogeneous personal histories.

Consequently, fairness assessment should not be limited to the performance of the final AI model. The clinical labels themselves should also be examined before model development, particularly when they represent constructs such as cognitive impairment or functional dependency that may be influenced by both clinical and social factors. Group fairness approaches provide a framework for identifying systematic disparities between population groups and can therefore complement conventional assessments of clinical validity and predictive performance \cite{rajkomar2018fairness}. Recent work has proposed frameworks to assess discrimination risks arising from training data composition before deployment, showing that balanced subgroup representation alone does not guarantee stable or equitable model behaviour \cite{bilionis2026jamia}. Moreover, fairness properties may change when models are retrained or updated with new data, which calls for continuous monitoring across the lifecycle \cite{bilionis2026embc}. The present study addresses a complementary question at the earliest stage: whether the outcome labels themselves are distributed unequally across sociodemographic groups. This perspective is also consistent with current European AI governance, which explicitly requires high-risk AI systems to address data governance, including the examination of possible biases in training, validation and testing datasets and the adoption of measures to detect, prevent and mitigate them (Article~10(2)(f)--(g)) \cite{euaiact2024}.

In this study, we examine the earliest stage of the AI lifecycle by auditing the clinical labels that would serve as target variables for models of functional and cognitive status in a cohort of 2,407 nursing home residents. We analyse the distribution of Barthel Index, MMSE and GDS classifications across four SDoH-related characteristics: sex, educational attainment, lifetime work activity and marital status. Disparities are quantified using two group fairness metrics, Disparate Impact (DI) and Statistical Parity Difference (SPD), computed separately for each level of functional dependency, cognitive impairment and dementia. Because no model is trained, these metrics are applied to the observed labels rather than to model predictions, and are interpreted as indicators of label-level disparity that downstream models could inherit.

\section{Methodology}

\subsection{Study Design, Setting and Data Sources}

A rigorous fairness evaluation requires detailed knowledge of the population from which clinical assessments are derived. Therefore, we first characterized the demographic and clinical profile of all nursing home residents included in the analysis, as the composition of the sample directly influences the detection and interpretation of bias across subgroups. Understanding the structure, variability, and representativeness of the data is essential before applying fairness metrics, particularly in older institutionalized populations where age, comorbidity burden, and gender distribution may be uneven.

This retrospective observational study used electronic health records (EHRs) from 12 DomusVi nursing homes in the Basque Country, Spain, covering the period 2011--2021. The data were obtained within the OROIPEN project, a multicentre geriatric assessment
programme aimed at the early detection of cognitive decline and the delivery of cognitive
stimulation in nursing home residents. Data were pseudonymised before analysis, with the correspondence table retained by the data controller. The dataset comprised:
\begin{itemize}
\item \textbf{Social determinants of health (SDoH) variables}: sex, age, marital status, educational attainment and lifetime work activity.
\item \textbf{Three standardised clinical assessments of functional and cognitive status}: the Barthel Index, the Mini-Mental State Examination (MMSE) and the Global Deterioration Scale (GDS).
\end{itemize}

\subsection{Study Population}

The study population comprised residents of the 12 participating DomusVi nursing homes
with electronic health records covering the period 2011--2021. Each source table was
deduplicated independently, retaining for each resident the most recent entry: the
demographic table yielded 5,087 residents, and the assessment tables 4,636 (Barthel
Index), 4,406 (MMSE) and 2,431 (GDS). Inclusion required complete results across all
three instruments simultaneously, since the audit compares label distributions between
them; 2,669 residents did not meet this combined criterion. A further 11 were excluded
because they lacked meal-plan records, a variable retained in the cohort definition
although not used in the present analysis, yielding a final sample of 2,407 residents..

Sociodemographic information was incomplete for part of the sample. Each analysis was therefore conducted on the residents with available data for the corresponding attribute, without imputation: sex ($n=2{,}407$), marital status ($n=2{,}403$), educational attainment ($n=2{,}037$) and work activity ($n=1{,}953$). The denominators of all reported percentages refer to these attribute-specific analytic samples.

Residents had a median age of 89 years (interquartile range 82 to 93; range 26 to 108),
and 81.6\% were aged 80 years or over. A minority of 95 residents (3.9\%) were aged under
65; this subgroup was predominantly never partnered (71.6\%) and had a markedly lower
prevalence of organic psychoses (5.3\% versus 24.3\%), suggesting a different pathway into
residential care. 

\subsection{Data Harmonization and Coding}

Given the heterogeneity of real-world clinical datasets, harmonizing categorical variables is a critical step to ensure fair and comparable subgroup evaluations. Differences in educational systems, occupational categories, and marital status definitions can introduce artificial disparities if not standardized appropriately. To avoid this, all sociodemographic variables were carefully recoded following international classification frameworks, allowing consistency across analyses and reducing the risk of spurious bias introduced by inconsistent labeling or fragmented categories.
\begin{itemize}
\item \textbf{Educational Level}: Educational categories were simplified according to the principles of UNESCO's International Standard Classification of Education (ISCED) \cite{isced2011}, reducing the original ten recorded categories to three broad levels:
\begin{itemize}
\item Low education: no schooling, primary incomplete/complete (ISCED 0--1)

\item Middle education: lower/upper secondary (ISCED 2--3)

\item High education: post-secondary, tertiary, or advanced degrees (ISCED 4--8)

\end{itemize}

\end{itemize}

\begin{itemize}
\item \textbf{Work Activity}: the main occupation held over the working life, as recorded at admission. Occupational categories were harmonized following the
International Standard Classification of Occupations (ISCO-08) framework \cite{isco08},
consolidating the original 14 recorded categories into the eight groups below. The
consolidation was performed at the level of the ISCO-08 occupational groupings rather
than by assigning individual ISCO codes, since the source records did not capture job
function in sufficient detail for code-level assignment:

\begin{itemize}

\item Unpaid domestic work: not included in ISCO-08, as the classification covers
only remunerated occupations. Maintained as a separate socioeconomic category.

\item Unskilled workers: non-qualified workers, service workers and occupational-centre
attendees, corresponding to elementary and service occupations.

\item Skilled workers: qualified workers, foremen and agricultural workers.

\item Intermediate technicians: middle-level technical staff and other technical personnel.

\item Senior executives: directors, managers and liberal professionals.

\item Employers and self-employed workers with employees, and own-account workers
without employees, retained as two separate categories because the source records
did not allow their job function to be determined.

\item State security forces: armed forces personnel and state security corps.

\end{itemize}

\end{itemize}

\begin{itemize}
\item \textbf{Marital Status}: Categories were harmonized into three clinically and
socio\-demographically meaningful groups to ensure consistency across fairness metrics
and outcome stratification:

\begin{itemize}

\item Never partnered: individuals who had never been married or in a stable union.

\item Partnered: currently married or living in a stable domestic partnership.

\item Previously partnered: widowed, divorced, or separated individuals.

\end{itemize}

\end{itemize}

\subsection{Outcome Definitions}

Fairness analysis depends fundamentally on how both outcomes and comparison groups are defined. For clinical assessment tools such as the Barthel Index, differences in category thresholds or reference groups can significantly affect fairness metrics. Establishing transparent and theoretically grounded definitions of dependency categories and reference groups ensures that disparities identified in the analysis reflect meaningful inequities rather than artifacts of inconsistent grouping.

Each instrument was summarised into three ordered categories of severity, defined by the participating gerontologists to retain clinical interpretability while avoiding the sparse minority classes that the original five-level and seven-stage scales would produce in this population:
\begin{itemize}
\item \textbf{Barthel Index} (0--100, in 5-point increments) \cite{mahoney1965barthel,shah1989}: \textit{dependent} (0--55, comprising the conventional total, severe and moderate dependence bands), \textit{mildly dependent} (60--95) and \textit{independent} (100).
\item \textbf{MMSE} (0--30) \cite{folstein1975minimental}: \textit{severe cognitive impairment} (0--12), \textit{moderate cognitive impairment} (13--20) and \textit{without cognitive impairment} (21--30). Uncorrected cut-off points were applied; education-adjusted thresholds validated for the Spanish population \cite{lobo1999} were not used, and the implications of this are considered in the Discussion.
\item \textbf{GDS} (stages 1--7) \cite{reisberg1982gds}: \textit{with dementia} (stages 4--7), \textit{slight dementia} (stage 3) and \textit{without dementia} (stages 1--2).
\end{itemize}

For every attribute, the outcome of interest was the probability of belonging to each of these categories, examined separately. For interpretation, the least severe category of each instrument (independent, without cognitive impairment, without dementia) is referred to as the favourable outcome and the most severe category as the unfavourable outcome.

\subsection{Reference Groups}

For each attribute, the reference group was fixed \textit{a priori} to the category identified as historically most disadvantaged in this population, drawing on evidence of social gradients in ageing and cognitive decline \cite{majoka2021sdoh,livingston2020dementia}: women for sex \cite{mielke2022sexgender}, low education for educational level \cite{arora2024education,lojo2012actividad}, unpaid domestic work for work activity \cite{feldberg2024occupation}, and previously partnered residents for marital status \cite{sommerlad2018marriage,liu2019marital}. The reference group was held constant across all outcome categories of all three instruments, so that every reported ratio is computed against the same denominator.

Because the reference group is the disadvantaged rather than the privileged category, the direction of the ratio is the reciprocal of the convention used in employment-discrimination settings. A value of DI above unity therefore indicates that the comparison group is classified in that outcome category more often than the historically disadvantaged reference group, and a value below unity indicates the opposite. Disparities were interpreted in both directions, since being classified more often in a favourable category and being classified more often in an unfavourable category carry opposite meanings.

\subsection{Fairness Metrics}

Two group fairness metrics were applied, both widely used in algorithmic fairness and implemented in standard auditing toolkits \cite{feldman2015,bellamy2019}. Because no predictive model was trained, both metrics were applied to the observed outcome labels $Y$ rather than to model predictions $\hat{Y}$. For an attribute $A$, a comparison group $g$, the reference group $r$ and an outcome category $c$:

\subsubsection{Disparate Impact (DI)}
\[
\mathrm{DI}_c(g) = \frac{P(Y = c \mid A = g)}{P(Y = c \mid A = r)}
\]
Values were interpreted according to the ``four-fifths rule'' derived from U.S. Equal Employment Opportunity Commission guidance \cite{uniformguidelines1978} and its reciprocal, as implemented in fairness toolkits \cite{bellamy2019}:
\begin{itemize}
    \item $\mathrm{DI} < 0.80$ $\rightarrow$ underrepresentation of group $g$ in category $c$.
    \item $0.80 \leq \mathrm{DI} \leq 1.25$ $\rightarrow$ acceptable range.
    \item $\mathrm{DI} > 1.25$ $\rightarrow$ overrepresentation of group $g$ in category $c$.
\end{itemize}
These thresholds were developed for employment selection contexts; their transfer to clinical assessment labels is a convention adopted here for comparability with the fairness literature rather than a clinically validated criterion, and is addressed in the Discussion.

\subsubsection{Statistical Parity Difference (SPD)}
\[
\mathrm{SPD}_c(g) = P(Y = c \mid A = g) - P(Y = c \mid A = r)
\]
SPD reflects the absolute difference in classification rates between groups; values close to zero indicate parity. The following threshold was defined \textit{a priori} as a descriptive heuristic and does not constitute a test of statistical significance:
\begin{itemize}
    \item $\lvert \mathrm{SPD} \rvert > 0.05$ $\rightarrow$ meaningful absolute difference.
\end{itemize}

Both metrics were computed separately for each outcome category of each instrument, allowing disparities to be located at specific levels of severity rather than assumed to be uniform across the population. The two metrics are reported together because DI is unstable when the rate in the reference group is low, as in the favourable categories of the Barthel Index, whereas SPD is less sensitive to low base rates but insensitive to large relative differences in rare categories. Normalising SPD by the rate in the reference group would yield $\mathrm{DI}-1$ and would
therefore add no information beyond the two metrics already reported; the pair is retained
because each is informative where the other is not.

\subsection{Statistical Analysis}

Associations between each sociodemographic attribute and the categories of each instrument were assessed using chi-square tests of independence. A global test was computed for each contingency table, and an additional test was computed for each outcome category against all others; both are reported. Statistical significance was set at $p<0.05$. No correction for multiple comparisons was applied; given the number of comparisons reported, nominal $p$ values and threshold-based alerts should be read as descriptive rather than confirmatory, and this is addressed in the Discussion.

Ninety-five per cent confidence intervals for DI and SPD were estimated by non-parametric bootstrap with 2,000 resamples of the attribute-specific analytic sample, a number that
stabilises the 2.5th and 97.5th percentiles without appreciable computational cost. Groups with fewer than 10 residents are reported for completeness but are not interpreted, since ratio-based metrics are unstable at such cell sizes. Analyses were conducted in Python 3.9 using \texttt{pandas}, \texttt{numpy} and \texttt{scipy}.

\section{Results}

This section presents the findings of the label audit. DI and SPD were computed for the Mini-Mental State Examination, the Barthel Index and the Global Deterioration Scale across the four social determinants of health, with bootstrap confidence intervals. For each instrument and attribute we first report the distribution of labels and then the fairness metrics. Results are presented separately by instrument to allow comparison of disparity patterns across assessment tools.

\subsection{Mini-Mental State Examination}

\subsubsection{Sex}

Table~\ref{tab:mmse_sex_dist} shows the distribution of cognitive impairment across sex. Women constitute 65.6\% of the analytic sample. Their share rises with severity, from 58.2\% of residents without cognitive impairment to 70.9\% of those with severe impairment. The difference is statistically significant for the extreme categories but not for moderate impairment ($p=0.369$).

\begin{table}[H]
\centering
\footnotesize
\setlength{\tabcolsep}{1.5pt}
\begin{tabular}{|p{1.9cm}|p{1.7cm}|p{1.7cm}|p{1.5cm}|p{1.1cm}|}
\hline
\multicolumn{5}{|c|}{\textbf{Mini-Mental Test by Sex}} \\
\hline
Imp. level & Men n (\%) & Women n (\%) & Total n (\%) & $p$ \\
\hline
Severe CI & 278 (29.1) & 676 (70.9) & 954 (100.0) & $<$0.001 \\
Moderate CI & 210 (32.9) & 428 (67.1) & 638 (100.0) & 0.369 \\
Without CI & 341 (41.8) & 474 (58.2) & 815 (100.0) & $<$0.001 \\
\hline
Total & 829 (34.4) & 1578 (65.6) & 2407 (100.0) & --- \\
\hline
\end{tabular}
\vspace{0.1cm}
\caption{Mini-Mental State Examination: distribution across sex. Values are $n$ (row \%); the Total row reports $n$ (\% of the analytic sample for this attribute, $N=2{,}407$). The $p$ column gives a Chi-square test for that outcome category against all others; global Chi-square $p<0.001$.}
\label{tab:mmse_sex_dist}
\end{table}

The fairness analysis (Table~\ref{tab:mmse_sex_fairness}) shows that men are classified as cognitively preserved more often than women (DI$=$1.369, 95\% CI 1.225 to 1.534; SPD$=+$0.111) and as severely impaired less often (DI$=$0.783, 95\% CI 0.700 to 0.875; SPD$=-$0.093). The moderate-impairment category is close to parity, with a confidence interval that includes unity.

\begin{table}[H]
\centering
\scriptsize
\setlength{\tabcolsep}{1.5pt}
\begin{tabular}{|p{1.5cm}|p{0.75cm}|p{1.45cm}|p{0.85cm}|p{1.6cm}|p{1.2cm}|}
\hline
\multicolumn{6}{|c|}{\textbf{Mini-Mental Test: Bias Alerts by Sex}} \\
\hline
Group & DI & 95\% CI & SPD & 95\% CI & Alert \\
\hline
\multicolumn{6}{|c|}{\textit{Severe CI}} \\
\hline
Women & 1.000 & Ref. & 0.000 & Ref. & Reference \\
Men & 0.783 & 0.70--0.87 & $-$0.093 & $-$0.13, $-$0.05 & $-$Bias, SPD \\
\hline
\multicolumn{6}{|c|}{\textit{Moderate CI}} \\
\hline
Women & 1.000 & Ref. & 0.000 & Ref. & Reference \\
Men & 0.934 & 0.81--1.08 & $-$0.018 & $-$0.05, 0.02 & No Sig. Diff. \\
\hline
\multicolumn{6}{|c|}{\textit{Without CI}} \\
\hline
Women & 1.000 & Ref. & 0.000 & Ref. & Reference \\
Men & 1.369 & 1.22--1.53 & $+$0.111 & 0.07, 0.15 & $+$Bias, SPD \\
\hline
\end{tabular}
\vspace{0.1cm}
\caption{Mini-Mental State Examination: fairness metrics (DI: Disparate Impact; SPD: Statistical Parity Difference) for sex. Reference group: women, fixed across all outcome categories. 95\% CI from 2{,}000 non-parametric bootstrap resamples. \textit{Notes:} $+$Bias $=$ DI $>$ 1.25; $-$Bias $=$ DI $<$ 0.80; SPD $=$ $|$SPD$| >$ 0.05; No Sig. Diff. $=$ no alert triggered.}
\label{tab:mmse_sex_fairness}
\end{table}

\subsubsection{Educational Attainment}

Table~\ref{tab:mmse_education_dist} presents the distribution across educational strata. Residents with low educational attainment constitute 87.6\% of the analytic sample for this attribute, reflecting the historical educational context of the generations represented in this sample. Their share is highest among residents with severe impairment (90.7\%) and lowest among those without impairment (83.5\%).

\begin{table}[H]
\centering
\footnotesize
\setlength{\tabcolsep}{1.5pt}
\begin{tabular}{|p{1.65cm}|p{1.3cm}|p{1.3cm}|p{1.3cm}|p{1.4cm}|p{1.0cm}|}
\hline
\multicolumn{6}{|c|}{\textbf{Mini-Mental Test by Education Level}} \\
\hline
Imp. level & Low n (\%) & Middle n (\%) & High n (\%) & Total n (\%) & $p$ \\
\hline
Severe CI & 733 (90.7) & 56 (6.9) & 19 (2.4) & 808 (100.0) & 0.002 \\
Moderate CI & 461 (88.3) & 49 (9.4) & 12 (2.3) & 522 (100.0) & 0.702 \\
Without CI & 590 (83.5) & 91 (12.9) & 26 (3.7) & 707 (100.0) & $<$0.001 \\
\hline
Total & 1784 (87.6) & 196 (9.6) & 57 (2.8) & 2037 (100.0) & --- \\
\hline
\end{tabular}
\vspace{0.1cm}
\caption{Mini-Mental State Examination: distribution across education level. Values are $n$ (row \%); the Total row reports $n$ (\% of the analytic sample for this attribute, $N=2{,}037$). The $p$ column gives a Chi-square test for that outcome category against all others; global Chi-square $p<0.001$.}
\label{tab:mmse_education_dist}
\end{table}

Table~\ref{tab:mmse_education_fairness} shows that residents with middle education are classified as cognitively preserved more often than the low-education reference group (DI$=$1.404, 95\% CI 1.184 to 1.631; SPD$=+$0.134) and as severely impaired less often (DI$=$0.695, 95\% CI 0.550 to 0.857; SPD$=-$0.125). Point estimates for high education follow the same direction, but all their confidence intervals include unity; with only 57 residents in this group, these estimates are imprecise and are not interpreted further.

\begin{table}[H]
\centering
\scriptsize
\setlength{\tabcolsep}{1.5pt}
\begin{tabular}{|p{1.5cm}|p{0.75cm}|p{1.45cm}|p{0.85cm}|p{1.6cm}|p{1.2cm}|}
\hline
\multicolumn{6}{|c|}{\textbf{Mini-Mental Test: Bias Alerts by Education Level}} \\
\hline
Group & DI & 95\% CI & SPD & 95\% CI & Alert \\
\hline
\multicolumn{6}{|c|}{\textit{Severe CI}} \\
\hline
Low & 1.000 & Ref. & 0.000 & Ref. & Reference \\
Middle & 0.695 & 0.55--0.86 & $-$0.125 & $-$0.19, $-$0.06 & $-$Bias, SPD \\
High & 0.811 & 0.53--1.13 & $-$0.078 & $-$0.20, 0.05 & SPD \\
\hline
\multicolumn{6}{|c|}{\textit{Moderate CI}} \\
\hline
Low & 1.000 & Ref. & 0.000 & Ref. & Reference \\
Middle & 0.967 & 0.74--1.23 & $-$0.008 & $-$0.07, 0.06 & No Sig. Diff. \\
High & 0.815 & 0.41--1.27 & $-$0.048 & $-$0.15, 0.07 & No Sig. Diff. \\
\hline
\multicolumn{6}{|c|}{\textit{Without CI}} \\
\hline
Low & 1.000 & Ref. & 0.000 & Ref. & Reference \\
Middle & 1.404 & 1.18--1.63 & $+$0.134 & 0.06, 0.20 & $+$Bias, SPD \\
High & 1.379 & 0.98--1.78 & $+$0.125 & $-$0.01, 0.26 & $+$Bias, SPD \\
\hline
\end{tabular}
\vspace{0.1cm}
\caption{Mini-Mental State Examination: fairness metrics (DI: Disparate Impact; SPD: Statistical Parity Difference) for education level. Reference group: low, fixed across all outcome categories. 95\% CI from 2{,}000 non-parametric bootstrap resamples. \textit{Notes:} $+$Bias $=$ DI $>$ 1.25; $-$Bias $=$ DI $<$ 0.80; SPD $=$ $|$SPD$| >$ 0.05; No Sig. Diff. $=$ no alert triggered.}
\label{tab:mmse_education_fairness}
\end{table}

\subsubsection{Work Activity}

Table~\ref{tab:mmse_occupation_dist} describes the distribution across occupational categories harmonised using ISCO-08. Unpaid domestic work is the largest category (59.8\%), followed by unskilled workers (20.2\%) and skilled workers (10.4\%). High-responsibility occupations together account for fewer than 6\% of residents. The global chi-square test is significant ($p=0.006$), driven by the without-impairment category ($p=0.001$).

\begin{table*}[!t]
\centering
\scriptsize
\setlength{\tabcolsep}{1pt}
\begin{tabular}{|p{1.7cm}|p{1.45cm}|p{1.45cm}|p{1.45cm}|p{1.45cm}|p{1.45cm}|p{1.45cm}|p{1.45cm}|p{1.45cm}|p{1.5cm}|p{0.95cm}|}
\hline
\multicolumn{11}{|c|}{\textbf{Mini-Mental Test by Work Activity}} \\
\hline
Imp. level & Domestic n (\%) & Unskilled n (\%) & Skilled n (\%) & Int. tech. n (\%) & Sr. execs n (\%) & Employers n (\%) & Self-emp. n (\%) & Security n (\%) & Total n (\%) & $p$ \\
\hline
Severe CI & 485 (61.8) & 160 (20.4) & 82 (10.4) & 19 (2.4) & 19 (2.4) & 6 (0.8) & 12 (1.5) & 2 (0.3) & 785 (100.0) & 0.181 \\
Moderate CI & 307 (61.3) & 98 (19.6) & 47 (9.4) & 11 (2.2) & 17 (3.4) & 6 (1.2) & 15 (3.0) & 0 (0.0) & 501 (100.0) & 0.094 \\
Without CI & 376 (56.4) & 136 (20.4) & 74 (11.1) & 36 (5.4) & 32 (4.8) & 6 (0.9) & 6 (0.9) & 1 (0.1) & 667 (100.0) & 0.001 \\
\hline
Total & 1168 (59.8) & 394 (20.2) & 203 (10.4) & 66 (3.4) & 68 (3.5) & 18 (0.9) & 33 (1.7) & 3 (0.2) & 1953 (100.0) & --- \\
\hline
\end{tabular}
\vspace{0.1cm}
\caption{Mini-Mental State Examination: distribution across work activity. Values are $n$ (row \%); the Total row reports $n$ (\% of the analytic sample for this attribute, $N=1{,}953$). The $p$ column gives a Chi-square test for that outcome category against all others; global Chi-square $p=0.006$.}
\label{tab:mmse_occupation_dist}
\end{table*}

Table~\ref{tab:mmse_occupation_fairness} reports fairness metrics against the unpaid domestic work reference group. Intermediate technicians (DI$=$1.694, 95\% CI 1.33 to 2.12) and senior executives (DI$=$1.462, 95\% CI 1.08 to 1.85) are classified as cognitively preserved more often, and correspondingly less often as severely impaired (DI$=$0.693 and DI$=$0.673, respectively). Self-employed residents without employees show the opposite pattern in the moderate category (DI$=$1.729, 95\% CI 1.05 to 2.40). Estimates for state security forces rest on three residents and are not interpreted.

\begin{table*}[!t]
\centering
\footnotesize
\setlength{\tabcolsep}{2pt}
\begin{tabular}{|p{3.5cm}|p{1.0cm}|p{2.0cm}|p{1.1cm}|p{2.2cm}|p{1.7cm}|}
\hline
\multicolumn{6}{|c|}{\textbf{Mini-Mental Test: Bias Alerts by Work Activity}} \\
\hline
Group & DI & 95\% CI & SPD & 95\% CI & Alert \\
\hline
\multicolumn{6}{|c|}{\textit{Severe CI}} \\
\hline
Unpaid domestic work & 1.000 & Ref. & 0.000 & Ref. & Reference \\
Unskilled workers & 0.978 & 0.85--1.12 & $-$0.009 & $-$0.07, 0.05 & No Sig. Diff. \\
Skilled workers & 0.973 & 0.80--1.15 & $-$0.011 & $-$0.09, 0.06 & No Sig. Diff. \\
Intermediate technicians & 0.693 & 0.45--0.97 & $-$0.127 & $-$0.23, $-$0.01 & $-$Bias, SPD \\
Senior executives & 0.673 & 0.43--0.96 & $-$0.136 & $-$0.24, $-$0.02 & $-$Bias, SPD \\
Employers/self-employed & 0.803 & 0.27--1.37 & $-$0.082 & $-$0.30, 0.15 & SPD \\
Self-employed (no emp.) & 0.876 & 0.48--1.29 & $-$0.052 & $-$0.22, 0.12 & SPD \\
State security forces$^{a}$ & 1.605 & 0.00--2.54 & $+$0.251 & $-$0.42, 0.61 & n.i. \\
\hline
\multicolumn{6}{|c|}{\textit{Moderate CI}} \\
\hline
Unpaid domestic work & 1.000 & Ref. & 0.000 & Ref. & Reference \\
Unskilled workers & 0.946 & 0.77--1.15 & $-$0.014 & $-$0.06, 0.04 & No Sig. Diff. \\
Skilled workers & 0.881 & 0.66--1.13 & $-$0.031 & $-$0.09, 0.03 & No Sig. Diff. \\
Intermediate technicians & 0.634 & 0.32--0.99 & $-$0.096 & $-$0.18, $-$0.00 & $-$Bias, SPD \\
Senior executives & 0.951 & 0.57--1.38 & $-$0.013 & $-$0.12, 0.09 & No Sig. Diff. \\
Employers/self-employed & 1.268 & 0.47--2.23 & $+$0.070 & $-$0.14, 0.32 & $+$Bias, SPD \\
Self-employed (no emp.) & 1.729 & 1.05--2.40 & $+$0.192 & 0.01, 0.36 & $+$Bias, SPD \\
State security forces$^{a}$ & 0.000 & 0.00--0.00 & $-$0.263 & $-$0.29, $-$0.24 & n.i. \\
\hline
\multicolumn{6}{|c|}{\textit{Without CI}} \\
\hline
Unpaid domestic work & 1.000 & Ref. & 0.000 & Ref. & Reference \\
Unskilled workers & 1.072 & 0.90--1.25 & $+$0.023 & $-$0.03, 0.08 & No Sig. Diff. \\
Skilled workers & 1.132 & 0.92--1.36 & $+$0.043 & $-$0.03, 0.11 & No Sig. Diff. \\
Intermediate technicians & 1.694 & 1.33--2.12 & $+$0.224 & 0.11, 0.35 & $+$Bias, SPD \\
Senior executives & 1.462 & 1.08--1.85 & $+$0.149 & 0.03, 0.27 & $+$Bias, SPD \\
Employers/self-employed & 1.035 & 0.33--1.78 & $+$0.011 & $-$0.22, 0.25 & No Sig. Diff. \\
Self-employed (no emp.) & 0.565 & 0.17--0.99 & $-$0.140 & $-$0.27, $-$0.00 & $-$Bias, SPD \\
State security forces$^{a}$ & 1.035 & 0.00--3.17 & $+$0.011 & $-$0.34, 0.68 & n.i. \\
\hline
\end{tabular}
\vspace{0.1cm}
\caption{Mini-Mental State Examination: fairness metrics (DI: Disparate Impact; SPD: Statistical Parity Difference) for work activity. Reference group: unpaid domestic work, fixed across all outcome categories. 95\% CI from 2{,}000 non-parametric bootstrap resamples. \textit{Notes:} $+$Bias $=$ DI $>$ 1.25; $-$Bias $=$ DI $<$ 0.80; SPD $=$ $|$SPD$| >$ 0.05; No Sig. Diff. $=$ no alert triggered. $^{a}$Group with fewer than 10 residents; estimates shown for completeness but not interpreted (n.i.).}
\label{tab:mmse_occupation_fairness}
\end{table*}

\subsubsection{Marital Status}

Table~\ref{tab:mmse_marital_dist} shows the distribution across marital status. Previously partnered residents are the largest category (58.1\%) and their share is stable across impairment levels. Never-partnered residents account for 16.9\% overall but 25.1\% of those without impairment, whereas partnered residents are concentrated in the severe category (31.4\%).

\begin{table}[H]
\centering
\scriptsize
\setlength{\tabcolsep}{1.5pt}
\begin{tabular}{|p{1.4cm}|p{1.3cm}|p{1.25cm}|p{1.25cm}|p{1.35cm}|p{0.95cm}|}
\hline
\multicolumn{6}{|c|}{\textbf{Mini-Mental Test by Marital Status}} \\
\hline
Imp. level & Prev. part. n (\%) & Partnered n (\%) & Never part. n (\%) & Total n (\%) & $p$ \\
\hline
Severe CI & 552 (58.0) & 299 (31.4) & 101 (10.6) & 952 (100.0) & $<$0.001 \\
Moderate CI & 384 (60.3) & 153 (24.0) & 100 (15.7) & 637 (100.0) & 0.401 \\
Without CI & 459 (56.4) & 151 (18.6) & 204 (25.1) & 814 (100.0) & $<$0.001 \\
\hline
Total & 1395 (58.1) & 603 (25.1) & 405 (16.9) & 2403 (100.0) & --- \\
\hline
\end{tabular}
\vspace{0.1cm}
\caption{Mini-Mental State Examination: distribution across marital status. Values are $n$ (row \%); the Total row reports $n$ (\% of the analytic sample for this attribute, $N=2{,}403$). The $p$ column gives a Chi-square test for that outcome category against all others; global Chi-square $p<0.001$.}
\label{tab:mmse_marital_dist}
\end{table}

Table~\ref{tab:mmse_marital_fairness} shows the most pronounced disparities for this instrument. Never-partnered residents are classified as cognitively preserved far more often than the previously partnered reference group (DI$=$1.531, 95\% CI 1.351 to 1.738; SPD$=+$0.175) and as severely impaired far less often (DI$=$0.630, 95\% CI 0.522 to 0.753; SPD$=-$0.146). Partnered residents show the reverse pattern in both categories. The moderate category is at parity.

\begin{table}[H]
\centering
\scriptsize
\setlength{\tabcolsep}{1.5pt}
\begin{tabular}{|p{1.5cm}|p{0.75cm}|p{1.45cm}|p{0.85cm}|p{1.6cm}|p{1.2cm}|}
\hline
\multicolumn{6}{|c|}{\textbf{Mini-Mental Test: Bias Alerts by Marital Status}} \\
\hline
Group & DI & 95\% CI & SPD & 95\% CI & Alert \\
\hline
\multicolumn{6}{|c|}{\textit{Severe CI}} \\
\hline
Prev. partnered & 1.000 & Ref. & 0.000 & Ref. & Reference \\
Partnered & 1.253 & 1.13--1.38 & $+$0.100 & 0.05, 0.15 & $+$Bias, SPD \\
Never partnered & 0.630 & 0.52--0.75 & $-$0.146 & $-$0.20, $-$0.09 & $-$Bias, SPD \\
\hline
\multicolumn{6}{|c|}{\textit{Moderate CI}} \\
\hline
Prev. partnered & 1.000 & Ref. & 0.000 & Ref. & Reference \\
Partnered & 0.922 & 0.78--1.07 & $-$0.022 & $-$0.06, 0.02 & No Sig. Diff. \\
Never partnered & 0.897 & 0.73--1.08 & $-$0.028 & $-$0.08, 0.02 & No Sig. Diff. \\
\hline
\multicolumn{6}{|c|}{\textit{Without CI}} \\
\hline
Prev. partnered & 1.000 & Ref. & 0.000 & Ref. & Reference \\
Partnered & 0.761 & 0.64--0.89 & $-$0.079 & $-$0.12, $-$0.04 & $-$Bias, SPD \\
Never partnered & 1.531 & 1.35--1.73 & $+$0.175 & 0.12, 0.23 & $+$Bias, SPD \\
\hline
\end{tabular}
\vspace{0.1cm}
\caption{Mini-Mental State Examination: fairness metrics (DI: Disparate Impact; SPD: Statistical Parity Difference) for marital status. Reference group: prev. partnered, fixed across all outcome categories. 95\% CI from 2{,}000 non-parametric bootstrap resamples. \textit{Notes:} $+$Bias $=$ DI $>$ 1.25; $-$Bias $=$ DI $<$ 0.80; SPD $=$ $|$SPD$| >$ 0.05; No Sig. Diff. $=$ no alert triggered.}
\label{tab:mmse_marital_fairness}
\end{table}

\subsection{Barthel Index}

\subsubsection{Sex}

Table~\ref{tab:bi_sex_dist} shows the distribution of functional dependency across sex. Women account for 68.3\% of dependent residents but only 45.5\% of independent residents, whereas men are comparatively more represented among the independent.

\begin{table}[H]
\centering
\footnotesize
\setlength{\tabcolsep}{1.5pt}
\begin{tabular}{|p{1.9cm}|p{1.7cm}|p{1.7cm}|p{1.5cm}|p{1.1cm}|}
\hline
\multicolumn{5}{|c|}{\textbf{Barthel Index by Sex}} \\
\hline
Dep. level & Men n (\%) & Women n (\%) & Total n (\%) & $p$ \\
\hline
Dependent & 476 (31.7) & 1027 (68.3) & 1503 (100.0) & $<$0.001 \\
Mild dependency & 298 (37.1) & 505 (62.9) & 803 (100.0) & 0.057 \\
Independent & 55 (54.5) & 46 (45.5) & 101 (100.0) & $<$0.001 \\
\hline
Total & 829 (34.4) & 1578 (65.6) & 2407 (100.0) & --- \\
\hline
\end{tabular}
\vspace{0.1cm}
\caption{Barthel Index: distribution across sex. Values are $n$ (row \%); the Total row reports $n$ (\% of the analytic sample for this attribute, $N=2{,}407$). The $p$ column gives a Chi-square test for that outcome category against all others; global Chi-square $p<0.001$.}
\label{tab:bi_sex_dist}
\end{table}

The fairness analysis (Table~\ref{tab:bi_sex_fairness}) indicates that men are classified as independent substantially more often than women (DI$=$2.276, 95\% CI 1.56 to 3.35), although the absolute difference is small (SPD$=+$0.037) because independence is rare in this population. Men are classified as dependent less often (DI$=$0.882, 95\% CI 0.823 to 0.944; SPD$=-$0.077). This divergence between a large ratio and a small absolute difference illustrates why both metrics are reported.

\begin{table}[H]
\centering
\scriptsize
\setlength{\tabcolsep}{1.5pt}
\begin{tabular}{|p{1.5cm}|p{0.75cm}|p{1.45cm}|p{0.85cm}|p{1.6cm}|p{1.2cm}|}
\hline
\multicolumn{6}{|c|}{\textbf{Barthel Index: Bias Alerts by Sex}} \\
\hline
Group & DI & 95\% CI & SPD & 95\% CI & Alert \\
\hline
\multicolumn{6}{|c|}{\textit{Dependent}} \\
\hline
Women & 1.000 & Ref. & 0.000 & Ref. & Reference \\
Men & 0.882 & 0.82--0.94 & $-$0.077 & $-$0.12, $-$0.04 & SPD \\
\hline
\multicolumn{6}{|c|}{\textit{Mild dependency}} \\
\hline
Women & 1.000 & Ref. & 0.000 & Ref. & Reference \\
Men & 1.123 & 1.00--1.25 & $+$0.039 & 0.00, 0.08 & No Sig. Diff. \\
\hline
\multicolumn{6}{|c|}{\textit{Independent}} \\
\hline
Women & 1.000 & Ref. & 0.000 & Ref. & Reference \\
Men & 2.276 & 1.56--3.35 & $+$0.037 & 0.02, 0.06 & $+$Bias \\
\hline
\end{tabular}
\vspace{0.1cm}
\caption{Barthel Index: fairness metrics (DI: Disparate Impact; SPD: Statistical Parity Difference) for sex. Reference group: women, fixed across all outcome categories. 95\% CI from 2{,}000 non-parametric bootstrap resamples. \textit{Notes:} $+$Bias $=$ DI $>$ 1.25; $-$Bias $=$ DI $<$ 0.80; SPD $=$ $|$SPD$| >$ 0.05; No Sig. Diff. $=$ no alert triggered.}
\label{tab:bi_sex_fairness}
\end{table}

\subsubsection{Educational Attainment}

Table~\ref{tab:bi_education_dist} shows the distribution of dependency levels across educational strata. The global chi-square test is significant ($p=0.019$), driven by the independent category ($p=0.011$); neither the dependent nor the mild dependency category reaches significance.

\begin{table}[H]
\centering
\footnotesize
\setlength{\tabcolsep}{1.5pt}
\begin{tabular}{|p{1.65cm}|p{1.3cm}|p{1.3cm}|p{1.3cm}|p{1.4cm}|p{1.0cm}|}
\hline
\multicolumn{6}{|c|}{\textbf{Barthel Index by Education Level}} \\
\hline
Dep. level & Low n (\%) & Middle n (\%) & High n (\%) & Total n (\%) & $p$ \\
\hline
Dependent & 1114 (88.8) & 105 (8.4) & 35 (2.8) & 1254 (100.0) & 0.053 \\
Mild dependency & 598 (86.4) & 74 (10.7) & 20 (2.9) & 692 (100.0) & 0.486 \\
Independent & 72 (79.1) & 17 (18.7) & 2 (2.2) & 91 (100.0) & 0.011 \\
\hline
Total & 1784 (87.6) & 196 (9.6) & 57 (2.8) & 2037 (100.0) & --- \\
\hline
\end{tabular}
\vspace{0.1cm}
\caption{Barthel Index: distribution across education level. Values are $n$ (row \%); the Total row reports $n$ (\% of the analytic sample for this attribute, $N=2{,}037$). The $p$ column gives a Chi-square test for that outcome category against all others; global Chi-square $p=0.019$.}
\label{tab:bi_education_dist}
\end{table}

Table~\ref{tab:bi_education_fairness} shows that residents with middle education are classified as independent more often than the low-education reference group (DI$=$2.149, 95\% CI 1.184 to 3.406), though the absolute difference is modest (SPD$=+$0.046) and does not exceed the 0.05 threshold. The same group is classified as dependent less often, with a significant absolute difference (SPD$=-$0.089, 95\% CI $-$0.162 to $-$0.018). Estimates for high education are imprecise throughout and are not interpreted.

\begin{table}[H]
\centering
\scriptsize
\setlength{\tabcolsep}{1.5pt}
\begin{tabular}{|p{1.5cm}|p{0.75cm}|p{1.45cm}|p{0.85cm}|p{1.6cm}|p{1.2cm}|}
\hline
\multicolumn{6}{|c|}{\textbf{Barthel Index: Bias Alerts by Education Level}} \\
\hline
Group & DI & 95\% CI & SPD & 95\% CI & Alert \\
\hline
\multicolumn{6}{|c|}{\textit{Dependent}} \\
\hline
Low & 1.000 & Ref. & 0.000 & Ref. & Reference \\
Middle & 0.858 & 0.74--0.97 & $-$0.089 & $-$0.16, $-$0.02 & SPD \\
High & 0.983 & 0.78--1.19 & $-$0.010 & $-$0.14, 0.12 & No Sig. Diff. \\
\hline
\multicolumn{6}{|c|}{\textit{Mild dependency}} \\
\hline
Low & 1.000 & Ref. & 0.000 & Ref. & Reference \\
Middle & 1.126 & 0.92--1.36 & $+$0.042 & $-$0.03, 0.12 & No Sig. Diff. \\
High & 1.047 & 0.68--1.42 & $+$0.016 & $-$0.11, 0.14 & No Sig. Diff. \\
\hline
\multicolumn{6}{|c|}{\textit{Independent}} \\
\hline
Low & 1.000 & Ref. & 0.000 & Ref. & Reference \\
Middle & 2.149 & 1.18--3.41 & $+$0.046 & 0.01, 0.09 & $+$Bias \\
High & 0.869 & 0.00--2.30 & $-$0.005 & $-$0.04, 0.05 & No Sig. Diff. \\
\hline
\end{tabular}
\vspace{0.1cm}
\caption{Barthel Index: fairness metrics (DI: Disparate Impact; SPD: Statistical Parity Difference) for education level. Reference group: low, fixed across all outcome categories. 95\% CI from 2{,}000 non-parametric bootstrap resamples. \textit{Notes:} $+$Bias $=$ DI $>$ 1.25; $-$Bias $=$ DI $<$ 0.80; SPD $=$ $|$SPD$| >$ 0.05; No Sig. Diff. $=$ no alert triggered.}
\label{tab:bi_education_fairness}
\end{table}

\subsubsection{Work Activity}

Table~\ref{tab:bi_occupation_dist} presents the distribution of dependency levels across occupational categories. No association was detected between work activity and dependency level, either globally ($p=0.821$) or for any individual category.

\begin{table*}[!t]
\centering
\scriptsize
\setlength{\tabcolsep}{1pt}
\begin{tabular}{|p{1.7cm}|p{1.45cm}|p{1.45cm}|p{1.45cm}|p{1.45cm}|p{1.45cm}|p{1.45cm}|p{1.45cm}|p{1.45cm}|p{1.5cm}|p{0.95cm}|}
\hline
\multicolumn{11}{|c|}{\textbf{Barthel Index by Work Activity}} \\
\hline
Dep. level & Domestic n (\%) & Unskilled n (\%) & Skilled n (\%) & Int. tech. n (\%) & Sr. execs n (\%) & Employers n (\%) & Self-emp. n (\%) & Security n (\%) & Total n (\%) & $p$ \\
\hline
Dependent & 746 (61.3) & 234 (19.2) & 127 (10.4) & 34 (2.8) & 41 (3.4) & 11 (0.9) & 21 (1.7) & 2 (0.2) & 1216 (100.0) & 0.529 \\
Mild dependency & 379 (57.4) & 143 (21.7) & 68 (10.3) & 30 (4.5) & 22 (3.3) & 6 (0.9) & 11 (1.7) & 1 (0.2) & 660 (100.0) & 0.524 \\
Independent & 43 (55.8) & 17 (22.1) & 8 (10.4) & 2 (2.6) & 5 (6.5) & 1 (1.3) & 1 (1.3) & 0 (0.0) & 77 (100.0) & 0.893 \\
\hline
Total & 1168 (59.8) & 394 (20.2) & 203 (10.4) & 66 (3.4) & 68 (3.5) & 18 (0.9) & 33 (1.7) & 3 (0.2) & 1953 (100.0) & --- \\
\hline
\end{tabular}
\vspace{0.1cm}
\caption{Barthel Index: distribution across work activity. Values are $n$ (row \%); the Total row reports $n$ (\% of the analytic sample for this attribute, $N=1{,}953$). The $p$ column gives a Chi-square test for that outcome category against all others; global Chi-square $p=0.821$.}
\label{tab:bi_occupation_dist}
\end{table*}

The fairness assessment (Table~\ref{tab:bi_occupation_fairness}) is largely consistent with this descriptive finding. Most groups fall within the acceptable DI range, with confidence intervals that include unity. Two exceptions stand out. Intermediate technicians are classified as mildly dependent more often than the unpaid domestic work reference group (DI$=$1.401, 95\% CI 1.028 to 1.817; SPD$=+$0.130) and show the largest absolute reduction in the dependent category (SPD$=-$0.124). Senior executives and employers show elevated point estimates for independence (DI$=$1.997 and DI$=$1.509), but both confidence intervals include unity and the absolute differences are below 0.05, so these should not be read as established disparities.

\begin{table*}[!t]
\centering
\footnotesize
\setlength{\tabcolsep}{2pt}
\begin{tabular}{|p{3.5cm}|p{1.0cm}|p{2.0cm}|p{1.1cm}|p{2.2cm}|p{1.7cm}|}
\hline
\multicolumn{6}{|c|}{\textbf{Barthel Index: Bias Alerts by Work Activity}} \\
\hline
Group & DI & 95\% CI & SPD & 95\% CI & Alert \\
\hline
\multicolumn{6}{|c|}{\textit{Dependent}} \\
\hline
Unpaid domestic work & 1.000 & Ref. & 0.000 & Ref. & Reference \\
Unskilled workers & 0.930 & 0.85--1.02 & $-$0.045 & $-$0.10, 0.01 & No Sig. Diff. \\
Skilled workers & 0.980 & 0.87--1.09 & $-$0.013 & $-$0.09, 0.06 & No Sig. Diff. \\
Intermediate technicians & 0.807 & 0.62--1.00 & $-$0.124 & $-$0.24, 0.00 & SPD \\
Senior executives & 0.944 & 0.75--1.13 & $-$0.036 & $-$0.16, 0.08 & No Sig. Diff. \\
Employers/self-employed & 0.957 & 0.57--1.31 & $-$0.028 & $-$0.28, 0.20 & No Sig. Diff. \\
Self-employed (no emp.) & 0.996 & 0.72--1.25 & $-$0.002 & $-$0.18, 0.16 & No Sig. Diff. \\
State security forces$^{a}$ & 1.044 & 0.00--1.62 & $+$0.028 & $-$0.64, 0.38 & n.i. \\
\hline
\multicolumn{6}{|c|}{\textit{Mild dependency}} \\
\hline
Unpaid domestic work & 1.000 & Ref. & 0.000 & Ref. & Reference \\
Unskilled workers & 1.119 & 0.95--1.30 & $+$0.038 & $-$0.02, 0.09 & No Sig. Diff. \\
Skilled workers & 1.032 & 0.84--1.25 & $+$0.010 & $-$0.05, 0.08 & No Sig. Diff. \\
Intermediate technicians & 1.401 & 1.03--1.82 & $+$0.130 & 0.01, 0.26 & $+$Bias, SPD \\
Senior executives & 0.997 & 0.66--1.35 & $-$0.001 & $-$0.11, 0.11 & No Sig. Diff. \\
Employers/self-employed & 1.027 & 0.38--1.71 & $+$0.009 & $-$0.20, 0.23 & No Sig. Diff. \\
Self-employed (no emp.) & 1.027 & 0.56--1.57 & $+$0.009 & $-$0.15, 0.18 & No Sig. Diff. \\
State security forces$^{a}$ & 1.027 & 0.00--3.17 & $+$0.009 & $-$0.34, 0.68 & n.i. \\
\hline
\multicolumn{6}{|c|}{\textit{Independent}} \\
\hline
Unpaid domestic work & 1.000 & Ref. & 0.000 & Ref. & Reference \\
Unskilled workers & 1.172 & 0.62--2.00 & $+$0.006 & $-$0.02, 0.03 & No Sig. Diff. \\
Skilled workers & 1.070 & 0.38--1.98 & $+$0.003 & $-$0.03, 0.03 & No Sig. Diff. \\
Intermediate technicians & 0.823 & 0.00--2.20 & $-$0.007 & $-$0.04, 0.04 & No Sig. Diff. \\
Senior executives & 1.997 & 0.42--4.00 & $+$0.037 & $-$0.02, 0.10 & $+$Bias \\
Employers/self-employed & 1.509 & 0.00--5.30 & $+$0.019 & $-$0.05, 0.15 & $+$Bias \\
Self-employed (no emp.) & 0.823 & 0.00--2.95 & $-$0.007 & $-$0.05, 0.07 & No Sig. Diff. \\
State security forces$^{a}$ & 0.000 & 0.00--0.00 & $-$0.037 & $-$0.05, $-$0.03 & n.i. \\
\hline
\end{tabular}
\vspace{0.1cm}
\caption{Barthel Index: fairness metrics (DI: Disparate Impact; SPD: Statistical Parity Difference) for work activity. Reference group: unpaid domestic work, fixed across all outcome categories. 95\% CI from 2{,}000 non-parametric bootstrap resamples. \textit{Notes:} $+$Bias $=$ DI $>$ 1.25; $-$Bias $=$ DI $<$ 0.80; SPD $=$ $|$SPD$| >$ 0.05; No Sig. Diff. $=$ no alert triggered. $^{a}$Group with fewer than 10 residents; estimates shown for completeness but not interpreted (n.i.).}
\label{tab:bi_occupation_fairness}
\end{table*}

\subsubsection{Marital Status}

Table~\ref{tab:bi_marital_dist} shows significant differences in dependency levels across marital status for every outcome category. Never-partnered residents make up 16.9\% of the analytic sample but 41.6\% of independent residents.

\begin{table}[H]
\centering
\scriptsize
\setlength{\tabcolsep}{1.5pt}
\begin{tabular}{|p{1.4cm}|p{1.3cm}|p{1.25cm}|p{1.25cm}|p{1.35cm}|p{0.95cm}|}
\hline
\multicolumn{6}{|c|}{\textbf{Barthel Index by Marital Status}} \\
\hline
Dep. level & Prev. part. n (\%) & Partnered n (\%) & Never part. n (\%) & Total n (\%) & $p$ \\
\hline
Dependent & 893 (59.6) & 401 (26.8) & 205 (13.7) & 1499 (100.0) & $<$0.001 \\
Mild dependency & 461 (57.4) & 184 (22.9) & 158 (19.7) & 803 (100.0) & 0.018 \\
Independent & 41 (40.6) & 18 (17.8) & 42 (41.6) & 101 (100.0) & $<$0.001 \\
\hline
Total & 1395 (58.1) & 603 (25.1) & 405 (16.9) & 2403 (100.0) & --- \\
\hline
\end{tabular}
\vspace{0.1cm}
\caption{Barthel Index: distribution across marital status. Values are $n$ (row \%); the Total row reports $n$ (\% of the analytic sample for this attribute, $N=2{,}403$). The $p$ column gives a Chi-square test for that outcome category against all others; global Chi-square $p<0.001$.}
\label{tab:bi_marital_dist}
\end{table}

Table~\ref{tab:bi_marital_fairness} shows the largest disparity observed in this study. Never-partnered residents are classified as independent more than three times as often as the previously partnered reference group (DI$=$3.528, 95\% CI 2.31 to 5.33; SPD$=+$0.074) and as dependent less often (DI$=$0.791, 95\% CI 0.711 to 0.874; SPD$=-$0.134). Partnered residents are close to parity across all three categories.

\begin{table}[H]
\centering
\scriptsize
\setlength{\tabcolsep}{1.5pt}
\begin{tabular}{|p{1.5cm}|p{0.75cm}|p{1.45cm}|p{0.85cm}|p{1.6cm}|p{1.2cm}|}
\hline
\multicolumn{6}{|c|}{\textbf{Barthel Index: Bias Alerts by Marital Status}} \\
\hline
Group & DI & 95\% CI & SPD & 95\% CI & Alert \\
\hline
\multicolumn{6}{|c|}{\textit{Dependent}} \\
\hline
Prev. partnered & 1.000 & Ref. & 0.000 & Ref. & Reference \\
Partnered & 1.039 & 0.97--1.11 & $+$0.025 & $-$0.02, 0.07 & No Sig. Diff. \\
Never partnered & 0.791 & 0.71--0.87 & $-$0.134 & $-$0.19, $-$0.08 & $-$Bias, SPD \\
\hline
\multicolumn{6}{|c|}{\textit{Mild dependency}} \\
\hline
Prev. partnered & 1.000 & Ref. & 0.000 & Ref. & Reference \\
Partnered & 0.923 & 0.80--1.06 & $-$0.025 & $-$0.07, 0.02 & No Sig. Diff. \\
Never partnered & 1.181 & 1.02--1.37 & $+$0.060 & 0.01, 0.12 & SPD \\
\hline
\multicolumn{6}{|c|}{\textit{Independent}} \\
\hline
Prev. partnered & 1.000 & Ref. & 0.000 & Ref. & Reference \\
Partnered & 1.016 & 0.55--1.64 & $+$0.000 & $-$0.02, 0.02 & No Sig. Diff. \\
Never partnered & 3.528 & 2.31--5.33 & $+$0.074 & 0.05, 0.11 & $+$Bias, SPD \\
\hline
\end{tabular}
\vspace{0.1cm}
\caption{Barthel Index: fairness metrics (DI: Disparate Impact; SPD: Statistical Parity Difference) for marital status. Reference group: prev. partnered, fixed across all outcome categories. 95\% CI from 2{,}000 non-parametric bootstrap resamples. \textit{Notes:} $+$Bias $=$ DI $>$ 1.25; $-$Bias $=$ DI $<$ 0.80; SPD $=$ $|$SPD$| >$ 0.05; No Sig. Diff. $=$ no alert triggered.}
\label{tab:bi_marital_fairness}
\end{table}

\subsection{Global Deterioration Scale}

\subsubsection{Sex}

Table~\ref{tab:gds_sex_dist} shows the distribution of dementia staging across sex. Women represent 69.2\% of residents staged with dementia, against 60.0\% of those without dementia and 59.1\% of those with slight dementia.

\begin{table}[H]
\centering
\footnotesize
\setlength{\tabcolsep}{1.5pt}
\begin{tabular}{|p{1.9cm}|p{1.7cm}|p{1.7cm}|p{1.5cm}|p{1.1cm}|}
\hline
\multicolumn{5}{|c|}{\textbf{GDS by Sex}} \\
\hline
Dementia level & Men n (\%) & Women n (\%) & Total n (\%) & $p$ \\
\hline
With dementia & 457 (30.8) & 1029 (69.2) & 1486 (100.0) & $<$0.001 \\
Slight dementia & 153 (40.9) & 221 (59.1) & 374 (100.0) & 0.005 \\
Without dementia & 219 (40.0) & 328 (60.0) & 547 (100.0) & 0.002 \\
\hline
Total & 829 (34.4) & 1578 (65.6) & 2407 (100.0) & --- \\
\hline
\end{tabular}
\vspace{0.1cm}
\caption{Global Deterioration Scale: distribution across sex. Values are $n$ (row \%); the Total row reports $n$ (\% of the analytic sample for this attribute, $N=2{,}407$). The $p$ column gives a Chi-square test for that outcome category against all others; global Chi-square $p<0.001$.}
\label{tab:gds_sex_dist}
\end{table}

Table~\ref{tab:gds_sex_fairness} shows that men are staged without dementia more often than women (DI$=$1.271, 95\% CI 1.09 to 1.48; SPD$=+$0.056) and with slight dementia more often (DI$=$1.318, 95\% CI 1.09 to 1.58), while being staged with dementia less often (DI$=$0.845, 95\% CI 0.787 to 0.906; SPD$=-$0.101). The direction is identical to that observed for the MMSE and the Barthel Index.

\begin{table}[H]
\centering
\scriptsize
\setlength{\tabcolsep}{1.5pt}
\begin{tabular}{|p{1.5cm}|p{0.75cm}|p{1.45cm}|p{0.85cm}|p{1.6cm}|p{1.2cm}|}
\hline
\multicolumn{6}{|c|}{\textbf{GDS: Bias Alerts by Sex}} \\
\hline
Group & DI & 95\% CI & SPD & 95\% CI & Alert \\
\hline
\multicolumn{6}{|c|}{\textit{With dementia}} \\
\hline
Women & 1.000 & Ref. & 0.000 & Ref. & Reference \\
Men & 0.845 & 0.79--0.91 & $-$0.101 & $-$0.14, $-$0.06 & SPD \\
\hline
\multicolumn{6}{|c|}{\textit{Slight dementia}} \\
\hline
Women & 1.000 & Ref. & 0.000 & Ref. & Reference \\
Men & 1.318 & 1.09--1.58 & $+$0.045 & 0.01, 0.08 & $+$Bias \\
\hline
\multicolumn{6}{|c|}{\textit{Without dementia}} \\
\hline
Women & 1.000 & Ref. & 0.000 & Ref. & Reference \\
Men & 1.271 & 1.09--1.48 & $+$0.056 & 0.02, 0.09 & $+$Bias, SPD \\
\hline
\end{tabular}
\vspace{0.1cm}
\caption{Global Deterioration Scale: fairness metrics (DI: Disparate Impact; SPD: Statistical Parity Difference) for sex. Reference group: women, fixed across all outcome categories. 95\% CI from 2{,}000 non-parametric bootstrap resamples. \textit{Notes:} $+$Bias $=$ DI $>$ 1.25; $-$Bias $=$ DI $<$ 0.80; SPD $=$ $|$SPD$| >$ 0.05; No Sig. Diff. $=$ no alert triggered.}
\label{tab:gds_sex_fairness}
\end{table}

\subsubsection{Educational Attainment}

Table~\ref{tab:gds_education_dist} presents the distribution of dementia staging across educational strata. Modest but statistically significant differences were observed globally ($p=0.015$), concentrated in the with-dementia ($p=0.016$) and without-dementia ($p=0.003$) categories.

\begin{table}[H]
\centering
\footnotesize
\setlength{\tabcolsep}{1.5pt}
\begin{tabular}{|p{1.65cm}|p{1.3cm}|p{1.3cm}|p{1.3cm}|p{1.4cm}|p{1.0cm}|}
\hline
\multicolumn{6}{|c|}{\textbf{GDS by Education Level}} \\
\hline
Dementia level & Low n (\%) & Middle n (\%) & High n (\%) & Total n (\%) & $p$ \\
\hline
With dementia & 1108 (89.1) & 101 (8.1) & 35 (2.8) & 1244 (100.0) & 0.016 \\
Slight dementia & 286 (88.0) & 31 (9.5) & 8 (2.5) & 325 (100.0) & 0.920 \\
Without dementia & 390 (83.3) & 64 (13.7) & 14 (3.0) & 468 (100.0) & 0.003 \\
\hline
Total & 1784 (87.6) & 196 (9.6) & 57 (2.8) & 2037 (100.0) & --- \\
\hline
\end{tabular}
\vspace{0.1cm}
\caption{Global Deterioration Scale: distribution across education level. Values are $n$ (row \%); the Total row reports $n$ (\% of the analytic sample for this attribute, $N=2{,}037$). The $p$ column gives a Chi-square test for that outcome category against all others; global Chi-square $p=0.015$.}
\label{tab:gds_education_dist}
\end{table}

Table~\ref{tab:gds_education_fairness} shows that residents with middle education are staged without dementia more often than the low-education reference group (DI$=$1.494, 95\% CI 1.18 to 1.83; SPD$=+$0.108) and with dementia less often (DI$=$0.830, 95\% CI 0.710 to 0.949; SPD$=-$0.106). High education shows no disparity whose confidence interval excludes unity.

\begin{table}[H]
\centering
\scriptsize
\setlength{\tabcolsep}{1.5pt}
\begin{tabular}{|p{1.5cm}|p{0.75cm}|p{1.45cm}|p{0.85cm}|p{1.6cm}|p{1.2cm}|}
\hline
\multicolumn{6}{|c|}{\textbf{GDS: Bias Alerts by Education Level}} \\
\hline
Group & DI & 95\% CI & SPD & 95\% CI & Alert \\
\hline
\multicolumn{6}{|c|}{\textit{With dementia}} \\
\hline
Low & 1.000 & Ref. & 0.000 & Ref. & Reference \\
Middle & 0.830 & 0.71--0.95 & $-$0.106 & $-$0.18, $-$0.03 & SPD \\
High & 0.989 & 0.76--1.20 & $-$0.007 & $-$0.15, 0.12 & No Sig. Diff. \\
\hline
\multicolumn{6}{|c|}{\textit{Slight dementia}} \\
\hline
Low & 1.000 & Ref. & 0.000 & Ref. & Reference \\
Middle & 0.987 & 0.66--1.34 & $-$0.002 & $-$0.06, 0.05 & No Sig. Diff. \\
High & 0.875 & 0.35--1.49 & $-$0.020 & $-$0.11, 0.08 & No Sig. Diff. \\
\hline
\multicolumn{6}{|c|}{\textit{Without dementia}} \\
\hline
Low & 1.000 & Ref. & 0.000 & Ref. & Reference \\
Middle & 1.494 & 1.18--1.83 & $+$0.108 & 0.04, 0.17 & $+$Bias, SPD \\
High & 1.124 & 0.62--1.67 & $+$0.027 & $-$0.08, 0.14 & No Sig. Diff. \\
\hline
\end{tabular}
\vspace{0.1cm}
\caption{Global Deterioration Scale: fairness metrics (DI: Disparate Impact; SPD: Statistical Parity Difference) for education level. Reference group: low, fixed across all outcome categories. 95\% CI from 2{,}000 non-parametric bootstrap resamples. \textit{Notes:} $+$Bias $=$ DI $>$ 1.25; $-$Bias $=$ DI $<$ 0.80; SPD $=$ $|$SPD$| >$ 0.05; No Sig. Diff. $=$ no alert triggered.}
\label{tab:gds_education_fairness}
\end{table}

\subsubsection{Work Activity}

Table~\ref{tab:gds_occupation_dist} describes the distribution of dementia staging across occupational categories. The global chi-square test is not significant ($p=0.134$), although the with-dementia category reaches nominal significance ($p=0.049$).

\begin{table*}[!t]
\centering
\scriptsize
\setlength{\tabcolsep}{1pt}
\begin{tabular}{|p{1.7cm}|p{1.45cm}|p{1.45cm}|p{1.45cm}|p{1.45cm}|p{1.45cm}|p{1.45cm}|p{1.45cm}|p{1.45cm}|p{1.5cm}|p{0.95cm}|}
\hline
\multicolumn{11}{|c|}{\textbf{GDS by Work Activity}} \\
\hline
Dementia level & Domestic n (\%) & Unskilled n (\%) & Skilled n (\%) & Int. tech. n (\%) & Sr. execs n (\%) & Employers n (\%) & Self-emp. n (\%) & Security n (\%) & Total n (\%) & $p$ \\
\hline
With dementia & 742 (61.7) & 229 (19.0) & 122 (10.1) & 30 (2.5) & 42 (3.5) & 11 (0.9) & 25 (2.1) & 2 (0.2) & 1203 (100.0) & 0.049 \\
Slight dementia & 170 (55.4) & 73 (23.8) & 37 (12.1) & 13 (4.2) & 10 (3.3) & 2 (0.7) & 1 (0.3) & 1 (0.3) & 307 (100.0) & 0.170 \\
Without dementia & 256 (57.8) & 92 (20.8) & 44 (9.9) & 23 (5.2) & 16 (3.6) & 5 (1.1) & 7 (1.6) & 0 (0.0) & 443 (100.0) & 0.389 \\
\hline
Total & 1168 (59.8) & 394 (20.2) & 203 (10.4) & 66 (3.4) & 68 (3.5) & 18 (0.9) & 33 (1.7) & 3 (0.2) & 1953 (100.0) & --- \\
\hline
\end{tabular}
\vspace{0.1cm}
\caption{Global Deterioration Scale: distribution across work activity. Values are $n$ (row \%); the Total row reports $n$ (\% of the analytic sample for this attribute, $N=1{,}953$). The $p$ column gives a Chi-square test for that outcome category against all others; global Chi-square $p=0.134$.}
\label{tab:gds_occupation_dist}
\end{table*}

Table~\ref{tab:gds_occupation_fairness} shows that intermediate technicians are staged with dementia less often than the unpaid domestic work reference group (DI$=$0.716, 95\% CI 0.523 to 0.917; SPD$=-$0.181) and without dementia more often (DI$=$1.590, 95\% CI 1.08 to 2.19; SPD$=+$0.129). Self-employed residents without employees are staged with slight dementia less often (DI$=$0.208, 95\% CI 0.00 to 0.71), an estimate resting on a single resident in that cell and therefore unstable despite the group exceeding the minimum cell size. The remaining groups fall within the acceptable range.

\begin{table*}[!t]
\centering
\footnotesize
\setlength{\tabcolsep}{2pt}
\begin{tabular}{|p{3.5cm}|p{1.0cm}|p{2.0cm}|p{1.1cm}|p{2.2cm}|p{1.7cm}|}
\hline
\multicolumn{6}{|c|}{\textbf{GDS: Bias Alerts by Work Activity}} \\
\hline
Group & DI & 95\% CI & SPD & 95\% CI & Alert \\
\hline
\multicolumn{6}{|c|}{\textit{With dementia}} \\
\hline
Unpaid domestic work & 1.000 & Ref. & 0.000 & Ref. & Reference \\
Unskilled workers & 0.915 & 0.83--1.00 & $-$0.054 & $-$0.11, 0.00 & SPD \\
Skilled workers & 0.946 & 0.84--1.07 & $-$0.034 & $-$0.11, 0.04 & No Sig. Diff. \\
Intermediate technicians & 0.716 & 0.52--0.92 & $-$0.181 & $-$0.30, $-$0.05 & $-$Bias, SPD \\
Senior executives & 0.972 & 0.79--1.16 & $-$0.018 & $-$0.14, 0.10 & No Sig. Diff. \\
Employers/self-employed & 0.962 & 0.59--1.31 & $-$0.024 & $-$0.26, 0.19 & No Sig. Diff. \\
Self-employed (no emp.) & 1.193 & 0.95--1.42 & $+$0.122 & $-$0.03, 0.26 & SPD \\
State security forces$^{a}$ & 1.049 & 0.00--1.63 & $+$0.031 & $-$0.64, 0.38 & n.i. \\
\hline
\multicolumn{6}{|c|}{\textit{Slight dementia}} \\
\hline
Unpaid domestic work & 1.000 & Ref. & 0.000 & Ref. & Reference \\
Unskilled workers & 1.273 & 0.97--1.60 & $+$0.040 & $-$0.00, 0.08 & $+$Bias \\
Skilled workers & 1.252 & 0.88--1.69 & $+$0.037 & $-$0.02, 0.10 & $+$Bias \\
Intermediate technicians & 1.353 & 0.66--2.09 & $+$0.051 & $-$0.05, 0.15 & $+$Bias, SPD \\
Senior executives & 1.010 & 0.47--1.64 & $+$0.002 & $-$0.08, 0.09 & No Sig. Diff. \\
Employers/self-employed & 0.763 & 0.00--1.99 & $-$0.034 & $-$0.16, 0.14 & $-$Bias \\
Self-employed (no emp.) & 0.208 & 0.00--0.71 & $-$0.115 & $-$0.16, $-$0.04 & $-$Bias, SPD \\
State security forces$^{a}$ & 2.290 & 0.00--7.06 & $+$0.188 & $-$0.16, 0.86 & n.i. \\
\hline
\multicolumn{6}{|c|}{\textit{Without dementia}} \\
\hline
Unpaid domestic work & 1.000 & Ref. & 0.000 & Ref. & Reference \\
Unskilled workers & 1.065 & 0.85--1.30 & $+$0.014 & $-$0.03, 0.06 & No Sig. Diff. \\
Skilled workers & 0.989 & 0.72--1.29 & $-$0.002 & $-$0.07, 0.06 & No Sig. Diff. \\
Intermediate technicians & 1.590 & 1.08--2.19 & $+$0.129 & 0.02, 0.25 & $+$Bias, SPD \\
Senior executives & 1.074 & 0.61--1.56 & $+$0.016 & $-$0.09, 0.12 & No Sig. Diff. \\
Employers/self-employed & 1.267 & 0.30--2.34 & $+$0.059 & $-$0.15, 0.29 & $+$Bias, SPD \\
Self-employed (no emp.) & 0.968 & 0.35--1.68 & $-$0.007 & $-$0.14, 0.14 & No Sig. Diff. \\
State security forces$^{a}$ & 0.000 & 0.00--0.00 & $-$0.219 & $-$0.24, $-$0.20 & n.i. \\
\hline
\end{tabular}
\vspace{0.1cm}
\caption{Global Deterioration Scale: fairness metrics (DI: Disparate Impact; SPD: Statistical Parity Difference) for work activity. Reference group: unpaid domestic work, fixed across all outcome categories. 95\% CI from 2{,}000 non-parametric bootstrap resamples. \textit{Notes:} $+$Bias $=$ DI $>$ 1.25; $-$Bias $=$ DI $<$ 0.80; SPD $=$ $|$SPD$| >$ 0.05; No Sig. Diff. $=$ no alert triggered. $^{a}$Group with fewer than 10 residents; estimates shown for completeness but not interpreted (n.i.).}
\label{tab:gds_occupation_fairness}
\end{table*}

\subsubsection{Marital Status}

Table~\ref{tab:gds_marital_dist} shows significant differences in dementia staging across marital status. Never-partnered residents account for 12.8\% of residents staged with dementia and 25.2\% of those staged without dementia.

\begin{table}[H]
\centering
\scriptsize
\setlength{\tabcolsep}{1.5pt}
\begin{tabular}{|p{1.4cm}|p{1.3cm}|p{1.25cm}|p{1.25cm}|p{1.35cm}|p{0.95cm}|}
\hline
\multicolumn{6}{|c|}{\textbf{GDS by Marital Status}} \\
\hline
Dementia level & Prev. part. n (\%) & Partnered n (\%) & Never part. n (\%) & Total n (\%) & $p$ \\
\hline
With dementia & 864 (58.3) & 428 (28.9) & 190 (12.8) & 1482 (100.0) & $<$0.001 \\
Slight dementia & 215 (57.5) & 82 (21.9) & 77 (20.6) & 374 (100.0) & 0.065 \\
Without dementia & 316 (57.8) & 93 (17.0) & 138 (25.2) & 547 (100.0) & $<$0.001 \\
\hline
Total & 1395 (58.1) & 603 (25.1) & 405 (16.9) & 2403 (100.0) & --- \\
\hline
\end{tabular}
\vspace{0.1cm}
\caption{Global Deterioration Scale: distribution across marital status. Values are $n$ (row \%); the Total row reports $n$ (\% of the analytic sample for this attribute, $N=2{,}403$). The $p$ column gives a Chi-square test for that outcome category against all others; global Chi-square $p<0.001$.}
\label{tab:gds_marital_dist}
\end{table}

Table~\ref{tab:gds_marital_fairness} shows that never-partnered residents are staged without dementia more often than the previously partnered reference group (DI$=$1.504, 95\% CI 1.268 to 1.773; SPD$=+$0.114) and with dementia less often (DI$=$0.757, 95\% CI 0.671 to 0.842; SPD$=-$0.150). Partnered residents show the reverse pattern. The slight dementia category is at parity.

\begin{table}[H]
\centering
\scriptsize
\setlength{\tabcolsep}{1.5pt}
\begin{tabular}{|p{1.5cm}|p{0.75cm}|p{1.45cm}|p{0.85cm}|p{1.6cm}|p{1.2cm}|}
\hline
\multicolumn{6}{|c|}{\textbf{GDS: Bias Alerts by Marital Status}} \\
\hline
Group & DI & 95\% CI & SPD & 95\% CI & Alert \\
\hline
\multicolumn{6}{|c|}{\textit{With dementia}} \\
\hline
Prev. partnered & 1.000 & Ref. & 0.000 & Ref. & Reference \\
Partnered & 1.146 & 1.07--1.22 & $+$0.090 & 0.05, 0.13 & SPD \\
Never partnered & 0.757 & 0.67--0.84 & $-$0.150 & $-$0.21, $-$0.10 & $-$Bias, SPD \\
\hline
\multicolumn{6}{|c|}{\textit{Slight dementia}} \\
\hline
Prev. partnered & 1.000 & Ref. & 0.000 & Ref. & Reference \\
Partnered & 0.882 & 0.68--1.11 & $-$0.018 & $-$0.05, 0.02 & No Sig. Diff. \\
Never partnered & 1.234 & 0.95--1.55 & $+$0.036 & $-$0.01, 0.08 & No Sig. Diff. \\
\hline
\multicolumn{6}{|c|}{\textit{Without dementia}} \\
\hline
Prev. partnered & 1.000 & Ref. & 0.000 & Ref. & Reference \\
Partnered & 0.681 & 0.55--0.83 & $-$0.072 & $-$0.11, $-$0.04 & $-$Bias, SPD \\
Never partnered & 1.504 & 1.27--1.77 & $+$0.114 & 0.06, 0.17 & $+$Bias, SPD \\
\hline
\end{tabular}
\vspace{0.1cm}
\caption{Global Deterioration Scale: fairness metrics (DI: Disparate Impact; SPD: Statistical Parity Difference) for marital status. Reference group: prev. partnered, fixed across all outcome categories. 95\% CI from 2{,}000 non-parametric bootstrap resamples. \textit{Notes:} $+$Bias $=$ DI $>$ 1.25; $-$Bias $=$ DI $<$ 0.80; SPD $=$ $|$SPD$| >$ 0.05; No Sig. Diff. $=$ no alert triggered.}
\label{tab:gds_marital_fairness}
\end{table}

\section{Discussion}

This study audited the clinical labels that would serve as target variables for AI models of functional and cognitive status in residential care, at the earliest stage of the AI lifecycle and before any model was trained. Across three instruments and four social determinants of health, the distribution of labels was not independent of residents' sociodemographic characteristics. The direction of these disparities was consistent: men, residents with middle educational attainment and never-partnered residents were classified in categories indicating preserved function or cognition more often than their respective reference groups, and in the most severe categories less often.

\subsection{Interpreting the Direction of Disparity}

Throughout this analysis, a DI above unity indicates that a group is classified in a given outcome category more often than the historically disadvantaged reference group. Because metrics were computed separately for every category, the same numerical value carries opposite substantive meaning depending on the category to which it refers: a value above unity in a favourable category indicates more frequent classification as preserved, whereas the same value in an unfavourable category indicates more frequent classification as impaired. Values are therefore not interpretable without reference to the category in which they arise.

\subsection{Main Findings}

Three patterns were robust in the sense that their confidence intervals excluded parity across more than one instrument.

First, sex-related disparities were present in all three instruments and in the same direction. Men were classified as independent (DI$=$2.276), without cognitive impairment (DI$=$1.369) and without dementia (DI$=$1.271) more often than women.

Second, marital status was associated with the largest disparity observed. Never-partnered residents were classified as independent on the Barthel Index more than three times as often as previously partnered residents (DI$=$3.528, 95\% CI 2.31 to 5.33), and showed the same direction on the MMSE (DI$=$1.531) and the GDS (DI$=$1.504).

Third, middle educational attainment was associated with more frequent classification in favourable categories across all three instruments (BI DI$=$2.149; MMSE DI$=$1.404; GDS DI$=$1.494). Estimates for high educational attainment pointed in the same direction but were imprecise throughout, as only 57 residents fell in that group, and are not interpreted. The pattern observed is therefore a step rather than a gradient: middle education differs from low education,
while high education cannot be distinguished from it. With 57 residents, this is as likely
to reflect limited precision as a genuine plateau in the educational effect.

By contrast, occupational gradients were weaker and less consistent than the descriptive pattern might suggest. For the Barthel Index, no association between work activity and dependency level was detected and most fairness estimates included parity; only intermediate technicians showed a disparity whose confidence interval excluded unity. For the MMSE and the GDS, intermediate technicians and senior executives were classified in favourable categories more often, but other occupational groups moved in opposite directions without an interpretable gradient. We therefore do not read these results as evidence of a monotonic socioeconomic gradient in assessment.

\subsection{Three Competing Explanations}

Observed label disparities are compatible with at least three distinct mechanisms, and the present design cannot distinguish between them. We set them out explicitly because the choice between them determines whether these disparities should be regarded as a data-quality problem, a measurement problem or a genuine epidemiological signal.

\emph{Genuine differences in health status.} Women live longer and, at any given age in advanced old age, are more likely to have accumulated functional limitations and dementia; sex differences in cognitive ageing are documented, though complex \cite{mielke2022sexgender}. Educational attainment is an established correlate of cognitive reserve, and a real protective association would produce exactly the pattern observed \cite{arora2024education,lojo2012actividad}. Under this reading, the labels are correct and the disparity reflects the epidemiology of the population.

\emph{Measurement properties of the instruments.} MMSE performance depends on schooling independently of underlying cognitive status, which is why education-adjusted cut-off points have been proposed, including for the Spanish population \cite{crum1993,lobo1999}. The thresholds applied here were uncorrected, so this explanation cannot be ruled out for the education-related findings. If uncorrected cut-off points were applied, part of the education-related disparity in MMSE labels would be measurement bias rather than a difference in cognition. A discriminant test is available in principle: the GDS is a clinician-staged instrument less dependent on literacy and schooling than the MMSE, so concentration of MMSE--GDS discordance in the low-education group would support a measurement explanation. We did not perform this analysis and identify it as a priority for future work.

\emph{Selective admission to residential care.} Admission to a nursing home is not independent of social circumstances. Residents with a surviving partner are typically admitted when home care has become unsustainable, and therefore at a more advanced stage of functional and cognitive decline; residents who never had a partner may be admitted earlier and in better condition for want of an informal carer. This selection mechanism would generate the marital-status pattern observed here without any difference in assessment practice, and it is the explanation we consider most plausible for the Barthel Index result. Distinguishing it from the alternatives requires information on the circumstances and timing of admission, which was not available.

A fourth possibility is that assessors apply the instruments differently depending on
residents' social characteristics. This cannot be excluded. It is tempting to read the
consistency across the three instruments as support for it, but that inference does not
hold: the MMSE and the GDS measure closely correlated constructs, so a genuine
epidemiological difference, or a shared selection mechanism, would produce concordant
results just as readily \cite{livingston2020dementia}. Cross-instrument agreement is
therefore compatible with all four explanations and discriminates between none of them.
Our data do not identify assessor behaviour as the operative mechanism, and earlier
framings of these results in terms of evaluator assumptions overstated what the design
can support.

\subsection{Implications for the AI Lifecycle}

The practical consequence of these findings does not depend on resolving the question of mechanism. A model trained on these labels would reproduce their distribution whatever its origin, because the labels define the target the model is fitted to. Where the disparity reflects genuine epidemiology, reproducing it may be clinically appropriate; where it reflects measurement properties or selection, reproducing it would transfer an artefact of the data-collection process into a decision-support system. The two cases are indistinguishable from the model's perspective and can only be separated upstream, at the stage audited here.

This argues for treating label auditing as a documented step of data governance rather than as a diagnostic that licenses automatic correction. Rebalancing or reweighting labels that encode a real health gradient would degrade clinical validity; leaving uncorrected a label set that encodes a measurement artefact would propagate inequity. The decision requires clinical and epidemiological judgement that fairness metrics alone cannot supply, and the appropriate output of an audit at this stage is therefore a characterisation of the disparities and of their candidate explanations, passed to the model-development stage as an explicit constraint.

Two further considerations follow. First, the audit is not a one-off exercise. Label distributions may shift over a ten-year period through changes in admission criteria, documentation practice or instrument versions, and fairness properties of models may change when they are retrained or updated on new data \cite{bilionis2026embc}. Repeating the audit at each data refresh is therefore preferable to performing it once. Second, label auditing is complementary to, rather than a substitute for, auditing the composition of the training data itself: representation balance in the input features does not guarantee stable or equitable model behaviour \cite{bilionis2026jamia}, and neither procedure detects the problems addressed by the other.

This positioning is consistent with the data-governance obligations that Article~10 of the EU AI Act places on high-risk systems, which require examination of possible biases in training, validation and testing datasets together with measures to detect, prevent and mitigate them \cite{euaiact2024}. An audit of the target variable, documented and repeated, is a concrete way of discharging that obligation at the point where the data enter the lifecycle.

\subsection{Methodological Contribution}

Group fairness metrics are normally applied to model outputs. Applying them to observed clinical labels, separately for each outcome category and with explicit confidence intervals, provides a reproducible characterisation of the disparities a downstream model would inherit. Two aspects of the implementation are worth noting. Stratifying by outcome category rather than dichotomising the instruments revealed that disparities are not uniform across severity: the moderate categories of the MMSE and the slight dementia category of the GDS were at parity on both metrics while the extreme categories were not, a pattern that a binary analysis would have concealed. Reporting DI together with SPD was likewise necessary, because the two diverge systematically in rare categories; independence on the Barthel Index, with a base rate of 4.2\%, produced large ratios alongside absolute differences below the threshold. Harmonisation through ISCED-2011 and ISCO-08 supports comparison with cohorts in other national contexts.

\subsection{Limitations}

Several limitations should be borne in mind when interpreting these findings. The analysis is bivariate throughout, and this bounds what can be separated. Occupation exhibits high collinearity/strong association with sex (Cramer's V = 0.53); notably, 84.76\% of residents who reported doing unpaid domestic work were women; the occupational and sex-related findings are
therefore best read together rather than as independent effects. For the same reason,
interactions between attributes---whether the education effect differs by sex, for
instance remain unexamined, despite the known intersectionality of these factors.
Confounder-adjusted regression, reporting odds ratios with confidence intervals as a
complementary analytical layer, would address both points and is the natural extension of this
work. A related caution applies to the number of comparisons: with DI and SPD reported for every
outcome category of three instruments across four attributes, some threshold crossings would be
expected by chance, and since no correction was applied the alerts are best read as descriptive
rather than confirmatory.

The data themselves impose further constraints. The design is cross-sectional: one assessment
per resident was retained and records were pooled across 2011--2021, so changes over that decade
in assessment practice, staff composition or instrument versions could not be examined, and may
contribute to the heterogeneity observed. Subgroup sizes are uneven, and the smallest groups
carry little information---high educational attainment ($n=57$) and the smaller occupational
categories yield confidence intervals too wide to interpret, while state security forces ($n=3$)
fell below the minimum cell-size threshold, so their estimates are reported for completeness
alone. Missing data were handled by complete-case analysis within each attribute, with
educational attainment missing for 370 residents and work activity for 454; since missingness in
EHR data is often non-random and may itself be associated with the social determinants under
study \cite{getzen2023missing}, its effect on the estimates cannot be signed.

Underlying all of this is a conceptual constraint. The fairness metrics applied here treat equal
rates across groups as the benchmark, which is not an appropriate normative standard wherever
genuine health differences exist. This is the principal interpretive limit of the approach, and
the reason the competing explanations set out above are presented as alternatives rather than
adjudicated between. The findings are also drawn from a single region of one country and from
three instruments, which limits how far they generalise.

\subsection{Conclusion}

Clinical labels used as target variables for AI in residential care were not distributed independently of residents' sex, educational attainment, work activity or marital status. The disparities were consistent in direction across three instruments, and the largest concerned never-partnered residents and the Barthel Index. Whether these disparities reflect genuine health differences, measurement properties of the instruments, assessment practice or selective admission cannot be determined from these data, and we do not claim to have identified assessor bias. What does follow is narrower and does not depend on that question: a model fitted to these labels would reproduce their distribution, and the distinction between a clinically appropriate gradient and a data-collection artefact can only be drawn before training. Auditing the target variable, documenting the candidate explanations and repeating the audit at each data refresh is therefore a practical first step of fairness assessment across the AI lifecycle, and a direct response to the data-governance obligations now placed on high-risk clinical AI.

\section*{Declarations}

\subsection*{Ethics approval and consent to participate}
Ethical approval was obtained from the Basque Country regional ethics committee
(ref. No PS2025047). In addition, written informed consent was obtained from study
participants. Data were pseudonymised prior to analysis, with the correspondence table
retained by the data controller; the authors had no access to directly identifying
information.

\subsection*{Funding}
This research was funded by the Government of the Basque Country under grant ZE-2021/00003.

\subsection*{CRediT authorship contribution statement}
\textbf{Isabel Amaya-Rodríguez:} Conceptualization, Methodology, Software, Formal analysis,
Investigation, Data curation, Visualization, Writing -- original draft.
\textbf{Nekane Larburu:} Project administration, Supervision, Writing.
\textbf{Cristina Martin:} Project administration, Supervision, Writing.
\textbf{Teresa Garcia-Navarro:} Formal analysis, Validation, Writing.
\textbf{Roberto González Novas:} Resources, Investigation, Validation.
\textbf{Alberto Rodríguez Taboada:} Resources, Investigation, Validation.
\textbf{Maite García-Sebastian:} Investigation, Validation.
\textbf{Ainara Estanga:} Investigation, Validation.
\textbf{Mirian Ecay-Torres:} Investigation, Validation.
\textbf{Pablo Martínez-Lage:} Investigation, Validation.
\textbf{Miguel A. Armengol de la Hoz:} Conceptualization, Methodology.
\textbf{Cristina Rubio-Escudero:} Conceptualization, Methodology, Supervision.
\textbf{Iván Macía:} Conceptualization, Methodology, Supervision,
Writing.

\subsection*{Data availability}
The individual-level clinical data analysed in this study are pseudonymised electronic
health records held by DomusVi and are subject to data protection restrictions under the
GDPR and Spanish Organic Law 3/2018; they cannot be made publicly available. All
contingency tables underlying the reported fairness metrics are included in the article.

\subsection*{Declaration of generative AI and AI-assisted technologies in the writing process}
During the preparation of this work the authors used generative AI tools to
assist with drafting and editing the English text. All analytical decisions, including the
choice of fairness metrics, reference groups and category thresholds, were made by the
authors, and all reported results were verified by the authors against the source data.
The authors reviewed and edited the content as needed and take full responsibility for the
content of the publication.


%\bibliography{references}

\end{document}